\documentclass[letters,fleqn,usenatbib]{mnras}

\usepackage{mathptmx}

\usepackage[T1]{fontenc}
\usepackage{ae,aecompl}

\usepackage{graphicx}	
\usepackage{amsmath}	
\usepackage{amssymb}	
\defcitealias{siudek18}{S1}

\title[Synergy between photo and spectro classifications]{The VIMOS Public Extragalactic Redshift Survey (VIPERS). 
   	Unsupervised classification with photometric redshifts: a method to accurately classify large galaxy samples without spectroscopic information
   	}

\author[M. Siudek et al.]{
M. Siudek$^{1,2,3}$\thanks{E-mail: msiudek@ifae.es},
K.~Ma{\l}ek$^{3}$,
A.~Pollo$^{3,4}$,
B.~R.~Granett$^{5,6}$,
M.~Scodeggio$^{7}$,
T.~Moutard$^{8,9}$, 
\newauthor 
A.~Iovino$^{5}$,
L.~Guzzo$^{6,5}$, 
B.~Garilli$^{7}$,        
M.~Bolzonella$^{10}$,    
S.~de la Torre$^{9}$,    
U.~Abbas$^{11}$,
\newauthor
C.~Adami$^{9}$,
D.~Bottini$^{7}$,
A.~Cappi$^{10,12}$,
O.~Cucciati$^{10}$,           
I.~Davidzon$^{9,10}$,   
P.~Franzetti$^{7}$,
\newauthor  
A.~Fritz$^{7}$,      
J.~Krywult$^{14}$,
V.~Le Brun$^{9}$,
O.~Le F\`evre$^{9}$,
D.~Maccagni$^{7}$,
F.~Marulli$^{13,15,10}$,
\newauthor
M.~Polletta$^{7,16,17}$,
L.A.M.~Tasca$^{9}$,
R.~Tojeiro$^{18}$, 
D.~Vergani$^{10}$,
A.~Zanichelli$^{19}$,
\newauthor 
S.~Arnouts$^{9,20}$,
J.~Bel$^{21}$,
E.~Branchini$^{22,23,24}$,
J.~Coupon$^{25}$,
G.~De Lucia$^{26}$,
O.~Ilbert$^{9}$,
\newauthor
L.~Moscardini$^{13,15,10}$,  
G.~Zamorani$^{10}$,
T.~T.~Takeuchi$^{27}$
\\\\
\textit{(Affiliations can be found after the references)}
}

\date{Accepted XXX. Received YYY; in original form ZZZ}

\pubyear{2015}

\begin{document}
\label{firstpage}
\pagerange{\pageref{firstpage}--\pageref{lastpage}}
\maketitle

\begin{abstract}
Techniques to classify galaxies solely based on photometry will be necessary for future large cosmology missions, such as Euclid or LSST.   
However, the precision of classification is always lower in photometric surveys and can be systematically biased with respect to classifications based upon spectroscopic data. 
We verified how precisely the detailed classification scheme introduced by~\citet[][hereafter: \citetalias{siudek18}]{siudek18} for galaxies at $z\sim0.7$ could be reproduced if only photometric data are available. 
We applied the Fisher Expectation-Maximization (FEM) unsupervised clustering algorithm to 54,293 VIPERS galaxies working in a parameter space of reliable photometric redshifts and 12 corresponding rest-frame magnitudes.
The FEM algorithm distinguishes four main groups: (1) red, (2) green, (3) blue, and (4) outliers.  
Each group is further divided into 3, 3, 4, and 2 subclasses, respectively. 
The accuracy of reproducing galaxy classes using spectroscopic data is high: $92\%$, $84\%$, $96\%$ for red, green, and blue classes, respectively, except for dusty star-forming galaxies. 
The presented verification of the photometric classification demonstrates that large photometric samples can be used to distinguish different galaxy classes at $z>0.5$ with an accuracy provided so far only by spectroscopic data except for particular galaxy classes. 
\end{abstract}

\begin{keywords}
galaxies: groups: general -- galaxies: evolution -- galaxies: star formation -- galaxies: stellar content
\end{keywords}



\section{Introduction}

Nowadays it has become imperative to apply efficient classification tools to process large astronomical datasets. 
Forthcoming surveys, such as the European Space Agency's Euclid mission~\citep{euclid2011} and the Large Synoptic Survey Telescope~\citep{ivezic2008,lsst2009} will collect data on an unprecedented scale and will benefit from advancements in machine learning and deep learning techniques (see~\citealp{FBD15} for a review). 
Machine learning methods have already proved to be successful in various astrophysical contexts, like discovering new exoplanets~\citep[e.g.][]{shallue2018}, characterise quasars~\citep[e.g.][]{parks2018}, distinguishing galaxy morphologies~\citep[e.g.][]{sanchez2018} or deriving photometric redshifts~\citep[e.g.][]{bilicki2018}. 

The upcoming datasets will allow unprecedented investigations into the physical processes underlying galaxy formation and evolution.  
A key step is to apply classification techniques to characterize the diversity of galaxy properties in order to infer the physical processes that drive the observed trends.  
Standard approaches to galaxy classification have relied on the bimodality in their colour distribution~\citep[e.g.][]{arnouts,fritz14,haines16,krywult,siudek17}. 
However, these approaches are incomplete in that they do not take advantage of all measurements available in the rich datasets.  
It has also been demonstrated that characterization based upon the bimodality can lead to systematic differences in the obtained properties depending on the selection criteria~\citep[e.g.][]{renzini,moresco13}. 

New techniques adopted from the field of machine learning provide an alternative approach to classify galaxies effectively.  
Among the machine learning tools, of particular interest are clustering algorithms which partition the sample into groups. 
Broadly speaking, they can be separated into two main classes: supervised (when the classifier is trained with a representative sample) and unsupervised methods (without any training process). 
While supervised algorithms have demonstrated reasonable efficiency in automatic classification of unknown sources if a good
training sample is provided~\citep[e.g.][]{Solarz2012,malek13,Krakowski16, Kurcz16,Solarz2017}, application of unsupervised machine learning techniques to galaxy classification is arguably yet to be proven. 
The advantage of unsupervised algorithms is that they can identify new classes of galaxies and interesting outliers in an objective manner.  
One of the challenges for unsupervised classifiers is the determination of the number of unique classes.  Algorithms can merge classes with diverse properties or subdivide the distribution into too many unphysical
classes and fail in following the continuous and physical nature of galaxy
types~\citep[e.g.][]{SDSS_class_2010, Hocking2018}. 

Unsupervised machine learning algorithms have so far been applied to classify spectroscopic galaxy measurements  based on Principal Component Analysis (PCA) and k-means analysis~\citep[e.g.][]{SDSS_class_2010,marchetti12,Peth2016}. 
Using photometric data for classification purposes is
less precise than with spectroscopic data due to larger errors. 
Unfortunately, obtaining full spectroscopic information in wide-field imaging surveys is not practical due to the large number of galaxies and high costs in observing time, especially for high-redshift sources. 
The next generation of extragalactic surveys will be dominated by multi-band photometry data. 
Therefore, there is a strong necessity to develop clustering tools able to provide galaxy classification using photometric redshift (hereafter $z_{phot}$) on a comparable accuracy level as when using spectroscopic redshift (hereafter $z_{spec}$). 

In~\citetalias{siudek18} we introduce a new approach to the galaxy classification based on the unsupervised Fisher Expectation-Maximization~\citep[hereafter FEM;][]{Bou2011} algorithm, which works in the multidimensional space of absolute magnitudes (in 12 different filters) and $z_{spec}$. 
The classification is thus based on a large parameter space, rather than just on the standard colour-colour plane, and therefore is more sensitive to the galaxy properties. 
The classification based upon broadband magnitudes allows detailed evolution studies of the red sequence, blue cloud and green valley. 
The FEM algorithm separates different galaxy classes and determines the optimal number of homogeneous classes automatically, yielding 11 galaxy classes from the earliest to the latest types with an additional group of outliers (mainly broad-line AGNs observed at $z\sim2$; \citetalias{siudek18}). 
The galaxy sequence is reflected in their colours, but also their morphological, physical and spectroscopic properties which are not included in the classification scheme.  
In this paper we present the photometric classification (hereafter PHOT classification) of VIPERS galaxies and its comparison to the $z_{spec}$-based classification (hereafter SPEC classification; \citetalias{siudek18}). 
The aim of the paper is to verify how precisely the SPEC classification can be reproduced by photometric data solely for a purpose of future surveys.

The paper is organized in the following way: 
In Sect. \ref{sec:data} we present the VIPERS data sample. 
Sect. \ref{sec:results} presents results and discusses their physical interpretation.
The summary is presented in Sect.~\ref{sec:summary}. 
Throughout this paper        
the cosmological framework assumes $\Omega_{m}$ = 0.30, $\Omega_{\Lambda}$ = 0.70, and $H_{0}=70$ $\rm{km s^{-1} Mpc^{-1}}$.

\section{Data and sample selection}\label{sec:data}

The analysis presented in this paper is based on the final galaxy sample from the VIMOS Public Extragalactic Redshift Survey~\citep[VIPERS,][]{scodeggio16} performed with the VIMOS spectrograph  mounted on the Very Large Telescope at ESO~\citep[VIMOS,][]{lefevre03}. 
VIPERS was designed to map the large-scale distribution of galaxies over an unprecedented volume of 5 x 10$^{-3}\it{h}^3\rm{Mpc^{-3}}$ in the redshift range  $0.5 < z < 1.2$.  
A simple and robust pre-selection in the ($u$-$g$) and ($r$-$i$) colour-colour plane was used to remove galaxies with $z < 0.5$~\citep{guzzo}. 
VIPERS provided spectroscopic redshifts of 86,775 galaxies limited to  $i_{AB}\leq22.5$ mag over an area of $\sim$23.5~$\deg^2$ and corresponding full photometrically-selected parent catalogue. 
The associated photometric catalogue consists of magnitudes based on $u$-, $g$-, $r$-, $i$- and $z$-band imaging from the CFHTLS T0007 release~\citep{Hudelot2012}, $FUV$ and $NUV$ observations from GALEX~\citep{Martin2005}, and new $Ks$-band observations conducted as part of the VIPERS Multi-Lambda Survey~\citep[VIPERS-MLS;][]{moutard16a}, complemented by $Ks$-band measurements from VIDEO~\citep{Jarvis2013}. 
A~detailed description of the VIPERS survey can be found in \cite{guzzo} and~\cite{scodeggio16}.

In this paper, we present the classification based on the absolute magnitudes and $z_{phot}$. 
Similarly as in the SPEC classification \citepalias{siudek18} the absolute magnitudes are derived by spectral energy distribution (SED) fitting described in~\cite{moutard16b}, except that now the computation is based on the $z_{phot}$ from the VIPERS-MLS~\citep[for full description of $z_{phot}$ measurements, please refer to][]{moutard16a}.

Our photometric sample includes the whole sample ({52,113} objects) classified with the SPEC classification (sample selection is discussed by \citetalias{siudek18}) completed by additional 2,180 sources. 
{2,148} sources of those 2,180 are spectroscopically confirmed stars, as such sources (on average $z_{phot}=0.7$) are expected to contaminate photometrically selected samples. 
The remaining 32 sources are high-redshift objects ($z_{spec}>1.5$) which are excluded from the spectroscopic sample defined in \citetalias{siudek18} as they are SED-fitting failures.
Our sources cover the redshift range $0.3 < z_{phot} <1.2$, reaching down to $z_{phot}\sim0$ and  up to $z_{phot}\sim4.8$.  

\begin{figure}
	\includegraphics[width=0.5\textwidth]{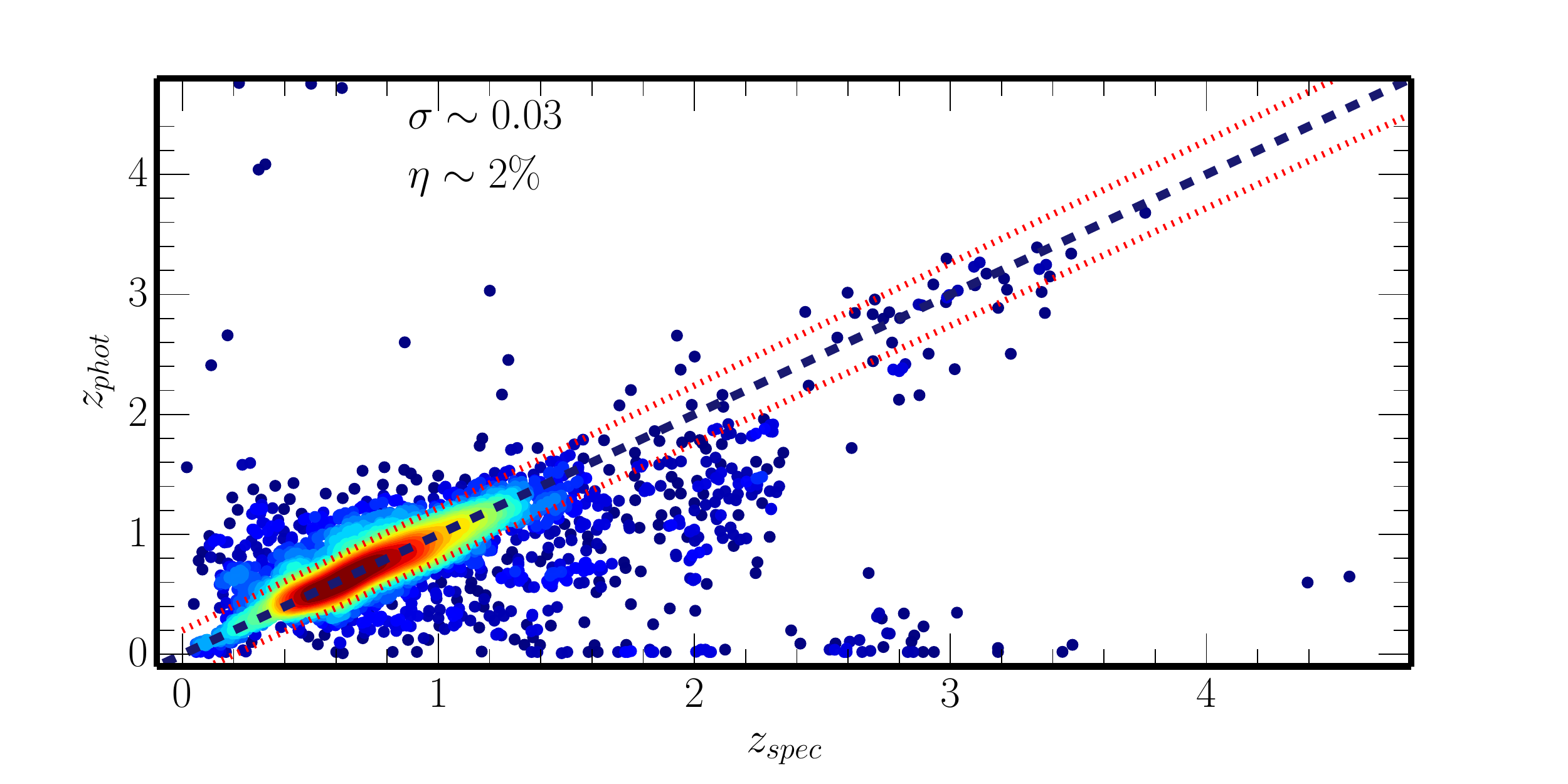}
	\caption{The comparison between $z_{phot}$ and the corresponding $z_{spec}$ of 54,293 VIPERS sources used in the following analysis. The blue dashed line corresponds to $z_{phot}=z_{spec}$. The red dotted lines show the threshold for defining outliers. }
	\label{fig:zphot}
\end{figure}

The comparison between $z_{phot}$ and $z_{spec}$ for the VIPERS sample is shown in Fig.~\ref{fig:zphot}. 
The measured scatter is $\sigma_{\delta z/(1+z)} \sim 0.03$ and the outlier rate\footnote{Following~\cite{ilbert2006} and~\cite{moutard16b} the scatter is defined as $\sigma_{\Delta z/(1+z)} = 1.48 \cdot median(|z_{spec}-z_{phot}|/(1+z_{spec}))$, and the photo-z outlier rate, $\eta$ as the percentage of galaxies with $|\Delta z|/(1+z) > 0.15$.} is $\eta \sim 2\%$.

\section{Results}\label{sec:results}

\subsection{The optimal class number}\label{sec:optimal_number}
To verify the applicability of this galaxy classification method in the case of $z_{phot}$ catalogues we applied for VIPERS photometric sample the same unsupervised algorithm as described in \cite{siudek18}. 
The parameter space was constructed from 12 absolute magnitudes (normalised rest-frame to $i$ band) and $z_{phot}$ of VIPERS galaxies~\cite[in contrast to $z_{spec}$ in][]{siudek18}. 
We ran the FEM algorithm on the PHOT sample based on the multidimensional space, and the optimal number of classes (12) and the best discriminative latent mixture (DLM) model (DBk) are established using the same approach as for SPEC classification \citepalias{siudek18}. 
We note that the optimal number of classes is the same as for SPEC classification; a new class of stars appears, but we are not able to distinguish one of the star-forming classes (namely, dusty star-forming galaxies from SPEC class 8 are merged with PHOT class~9; see Fig.~\ref{fig:nuvrrk} and~\ref{fig:sankey}) and so the number of classes is preserved. 
Both in the SPEC and PHOT classification process, the  sources are not strictly assigned to one class, but the probability of being a member of each class is given. 
However, $94\%$ of sources have their assignment well defined, i.e. have a high probability ($>75\%$) of belonging to one class. 
In the following discussion we use only class members with a high probability of belonging to a given class, rejecting 3,007 objects with low class membership probability ($<50\%$) or with a high ($>45\%$) probability of belonging to the second group (see \citetalias{siudek18} for details). 


\subsection{Galaxy class properties}\label{sec:nuvrk}

\begin{figure}
	\includegraphics[width=0.49\textwidth]{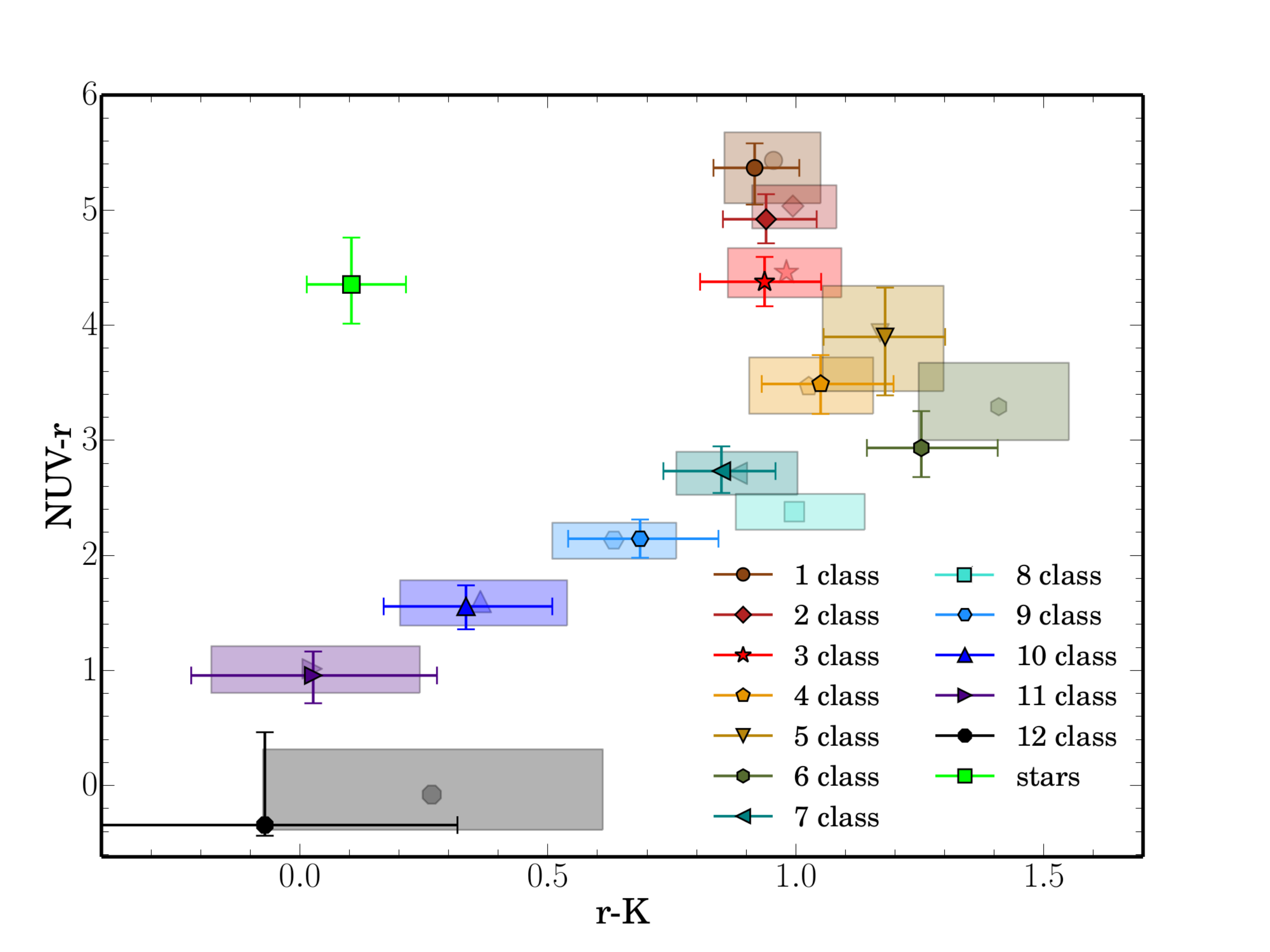}
	\caption{The $NUVrK$ diagram of the VIPERS galaxies classified into 12 classes according to the FEM algorithm using $z_{phot}$ (points with error bars) and $z_{spec}$ (rectangles). The error bars and size of squares correspond to the 25th, and 75th percentile range of the galaxy colour distribution. The median colours are marked with points. 
	}
	\label{fig:nuvrrk}
\end{figure}
The 12 different photometric classes (hereafter PHOT class) within the VIPERS sample follow the trend from red, passive, through green, intermediate, to blue, star-forming galaxy types, with an additional groups of outliers (stars and AGNs). 
Even if such a detailed classification is possible only in the full multidimensional parameter space, it is well reflected by the positions of groups on the $(NUV-r)$ vs. $(r-K)$ diagram~\cite[hereafter $NUVrK$,][]{arnouts} diagram\footnote{Note, however, that the three ($NUV$, $r$, $K$) parameter space would not be large enough to create such a fine division.}. 
Fig.~\ref{fig:nuvrrk} presents a comparison of the distribution of the FEM classes in the $NUVrK$ diagram, when $z_{phot}$ and $z_{spec}$ were used in the classification procedure. 


The galaxy sequence obtained from the PHOT classification almost exactly mirrors the spectroscopic classes (hereafter SPEC classes). 
Fig.~\ref{fig:nuvrrk} shows that the median $NUV-r$ and $r-K$ colours of red galaxies (classes 1--3) are similar in both classifications. 
Green galaxies (classes 4--6) are also located in the same area in the $NUVrK$ diagram in SPEC and PHOT classifications. 
The main difference is displayed by the group of star-forming galaxies. 
In the PHOT classification four star-forming galaxy classes are found, whereas in the SPEC classification five star-forming groups are distinguished. 
One additional class in SPEC classification (SPEC class 8) is characterized by redder $rK$ colours ($r-K_{median}=1.00^{+0.12}_{-0.14}$) and higher star formation rate (hereafter $SFR$; $SFR=1.41^{+0.28}_{-0.30}$) than the remaining star-forming classes. 
These galaxies are mostly assigned to the PHOT class 9 of star-forming galaxies ($\sim75\%$; see also Fig.~\ref{fig:sankey}).
This might be a consequence of the discrete nature of SED templates, which resulted in the lowest $z_{phot}$ accuracy ($\sigma_{\delta z /(1+z)} \sim 0.037$) for galaxies within SPEC class 8 among all classes (except class 12) and yields the inability of distinguishing  more dusty star-forming galaxies. 
We note that also in the SPEC classification those galaxies are not straightforwardly distinguished. 
Galaxies in SPEC class~8 are characterized by lower membership probability than other SPEC classes (see Fig.~C.1 in~\citetalias{siudek18}). 
Moreover, the SPEC class~8 is distinguished on the last cycle of the FEM algorithm, i.e. the class is absent if we consider a SPEC classification with only 11~classes (Fig.~B.2 and~B.3 in~\citetalias{siudek18}). 
Extremely dusty green galaxies (class~6) are, however, properly identified in the PHOT classification (see Fig.~\ref{fig:nuvrrk} and~\ref{fig:sankey}).

Furthermore, in the PHOT classification an additional group of outliers (marked with green in Fig.~\ref{fig:nuvrrk}) is separated. 
The class of outliers with red $NUV-r$ and blue $r-K$ colours is mainly ($96\%$) composed of stars (not included in the original SPEC sample but added to the PHOT sample to model photometric data in a realistic way).
More than half of the stars present in the photometric dataset (1,234 out of 2,148; $57\%$) are assigned to this class.  
The remaining 914 stars are distributed in different groups.
However, they constitute just a small fraction of all members of the other classes except for class 12 ($42\%$; 55 stars). 
Class 12, besides stars, consists of broad-line AGNs ($50\%$). 
In the case of the SPEC classification it is a pure sample of broad-line AGNs ($95\%$). 
%
%
%

\begin{figure}
	\includegraphics[width=0.5\textwidth]{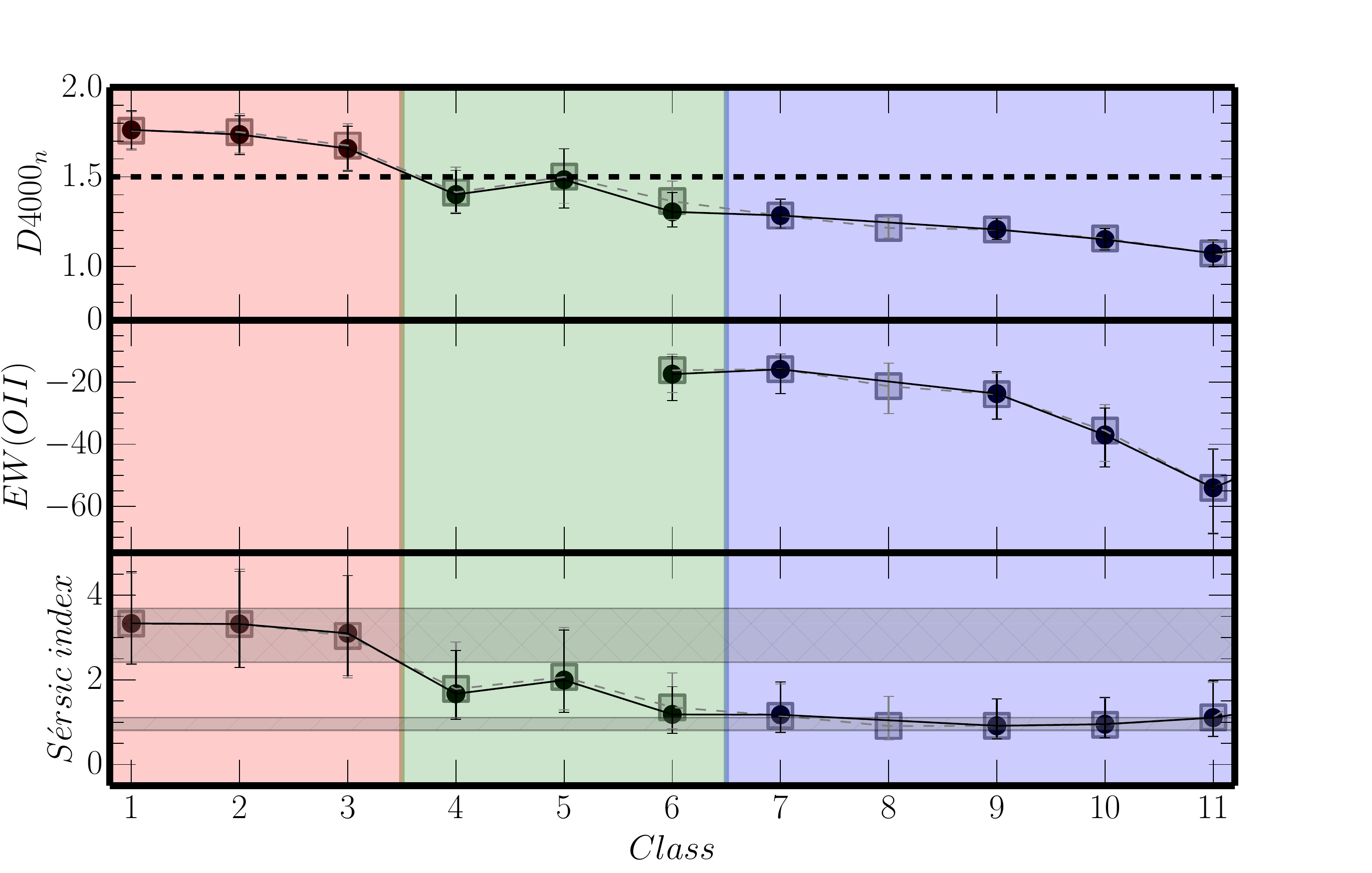}
	\caption{
		The distribution of the median values of $D4000_{n}$, $EW(OII)$, and S\'{e}rsic index for SPEC (marked with grey open squares) and PHOT (marked with black dots) classifications for red (classes 1--3), green (classes 4--6), star-forming (classes 7--11) galaxies 
		are shown in red, green, and blue, 
		respectively. The error bars correspond to 1st and 3rd quartile of the parameter distribution. The division between red passive and blue active galaxies based on $D4000_{n}$~\citep{kauffmann03} is marked with a black dashed line. The typical mean S\'{e}rsic index ranges for VIPERS red passive, and blue star-forming galaxies~\citep{krywult} are marked with grey shadow areas. The $[OII]\lambda3727$ line has not been detected in the majority of galaxies from SPEC and PHOT classes 1--5.  
	}
	\label{fig:properties}
\end{figure}

PHOT classification is able to recover the trends from SPEC classification in physically important parameters like line strength and morphology.
Fig.~\ref{fig:properties} presents the comparison between the spectroscopic and morphological galaxy class properties: the strength of the $4000\AA$ break~\citep[$D4000_{n}$, as defined by][]{balogh1999}, the equivalent widths  for $[OII]\lambda3727$, and the S\'{e}rsic index when $z_{phot}$ and $z_{spec}$ were used in the classification procedure are shown as a function of the class number. 
With the exception of SPEC class 8 (blue, dusty star-forming galaxies not distinguishable in the PHOT classification), the median properties of each class agree well in the two classifications and do not show any significant differences.  

\subsection{Statistical differences between spectroscopic and photometric classifications}\label{sec:flowchart}
\begin{figure}
	\includegraphics[width=0.49\textwidth]{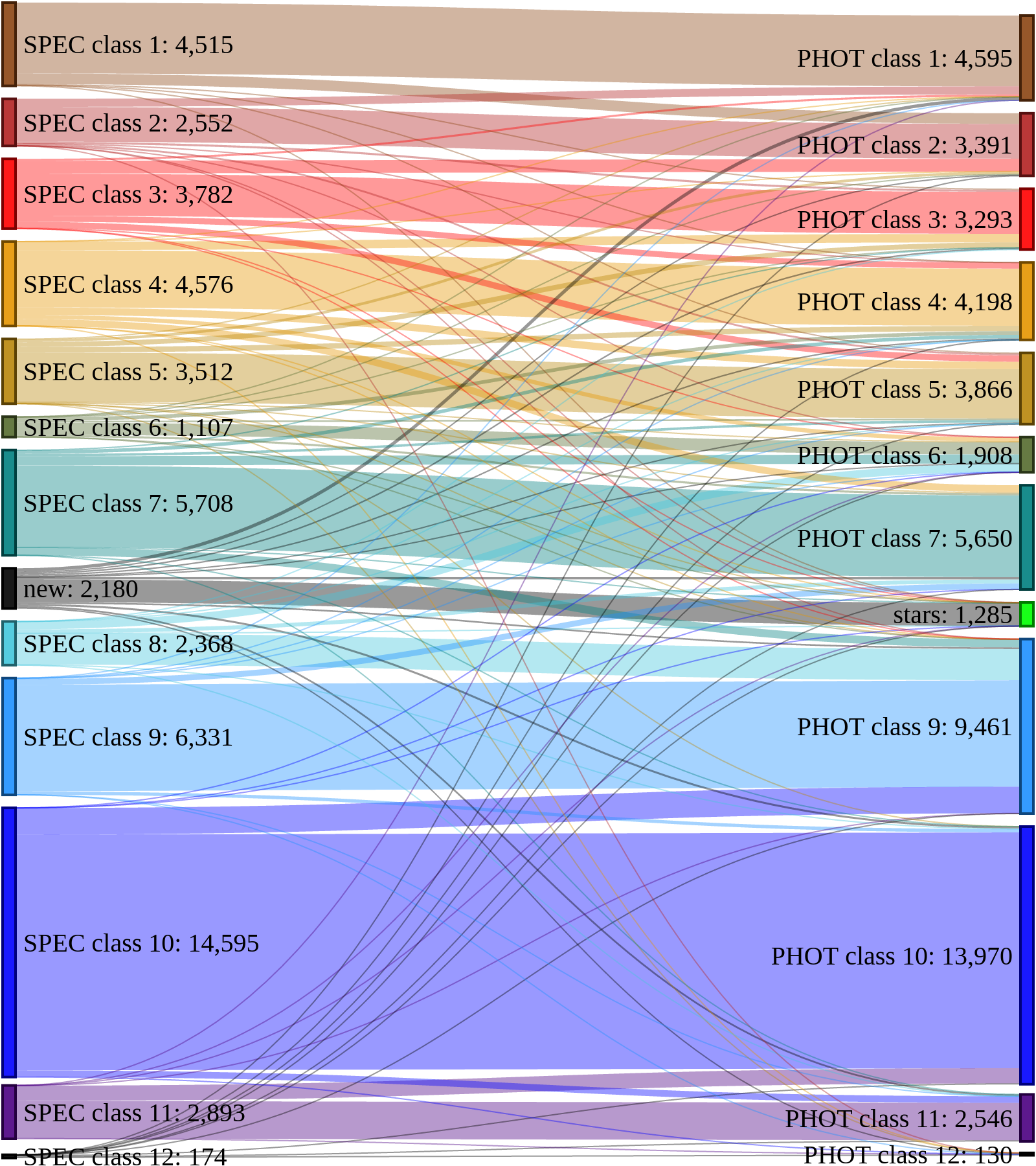}
	\caption{The distribution of 52,113 VIPERS galaxies and 2,180 additional sources (mainly stars) within 12 SPEC (left side) and PHOT classes (right side). The coloured lines show the correspondence between object assignments in the SPEC and PHOT classifications.   }
	\label{fig:sankey}
\end{figure}

Fig.~\ref{fig:sankey} presents the distribution of VIPERS sources within SPEC classes \citepalias{siudek18} and their migration to PHOT classes. 
The majority of objects ($>70\%$) are assigned to the same class independently of the fact that $z_{spec}$ or $z_{phot}$ is used. 
However, a small "exchange" between classes occurred, mainly between classes located close to each other in the multidimensional feature space. 
For example PHOT class 2 (marked with dark red in Fig.~\ref{fig:sankey}) is built mainly from galaxies assigned to SPEC class~2 ($\sim56\%$; dark red), but also includes objects from SPEC class~1 ($\sim17\%$; brown), and~3 ($\sim20\%$; red).
There are also more extreme "jumping-objects" where sources assigned to e.g., red and passive SPEC classes are classified as intermediate or star-forming PHOT classes, and this is seen as thin flows between distant groups in Fig.~\ref{fig:sankey}. 
However, jumps over more than 2 neighbouring classes are very rare ($\sim2\%$). 
Less secure assignment of these sources is also mirrored by their lower membership probability (mean probability for jumping-objects is lower by $\sim7\%$ than those for members of main streams). 
Therefore, some objects with a low PHOT membership probability do not migrate from the SPEC class to the corresponding PHOT class, but drastically change their classification. 
These galaxies are also characterized by lower $z_{phot}$ accuracy ($\sigma_{\Delta z/(1+z)} \sim 0.05$), which may explain the difficulties in the assignation of the proper class.

\section{Conclusions}\label{sec:summary}
We present an unsupervised classification of VIPERS sources observed at $z\sim0.7$ based on the $z_{phot}$ and corresponding absolute magnitudes.  
The unsupervised photometric classification does not require additional training with spectroscopic samples, after photometric redshifts have been estimated.

Three classes of red passive galaxies (classes 1--3) are correctly recovered if no spectroscopy information is provided (with an accuracy of $92\%$; see Tab.~\ref{table:eff}). 
Also green, intermediate galaxies (classes 4--6) are accurately ($84\%$) separated when only $z_{phot}$ is provided. 
Spectroscopy is also not necessary to separate blue, star-forming galaxies as a whole class, as the algorithm using $z_{phot}$ is able to almost perfectly identify the same group as SPEC classification ($96\%$). 
However, spectroscopic information is needed to obtain a more detailed picture of some of the blue subclasses. 
In particular, the PHOT classification is not able to separate dusty star-forming galaxies (SPEC class 8) 
from less dusty star-forming galaxies (class 9). 
This might be a consequence of the SED fitting procedure adapted to the VIPERS dataset~\citep{moutard16a}, which is not able to describe properly dusty star-forming galaxies. 
Possibly, a better set of templates might improve the outcome. 
In the PHOT sample, which includes stars, the algorithm, in some cases, is not able to distinguish stars and AGNs and, therefore, assigned them to the same class. 
This results in (expected) differences between PHOT class 12 and SPEC class 12. 
A detailed efficiency of the PHOT classification is given in Tab.~\ref{table:eff}. 

\begin{table}
	\centering                         
	\begin{tabular}{p{0.47cm} p{0.71cm} p{0.71cm} p{0.85cm}|p{0.65cm} p{0.71cm} p{0.71cm} p{0.85cm}}     
		\hline 
		Cls & $N_{spectro}$ & $N_{photo}$ & A [$\%$]& Cls & $N_{spectro}$ & $N_{photo}$ & A [$\%$]\\     
		\hline 
		\hline
		1 & 4,515 & 3,838 & 85 & 7 & 5,708 & 4,432 & 78 \\
		2 & 2,552 & 1,884 & 74 & 9 & 6,331 & 5,773 & 91\\
		3 & 3,782 & 2,279 & 60 & 10 & 14,595 & 12,781 & 88\\
		\cline{1-4}
		\textbf{1-3} & \textbf{10,849} & \textbf{9,983} & \textbf{92} & 11 & 2,893 & 2,036 & 70\\
		\cline{1-8}			
		4 & 4,576 & 3,100 & 68 & \textbf{7-11} & \textbf{31,896} & \textbf{30,552} & \textbf{96}\\
		\cline{5-8}
		5 & 3,512 & 2,673 & 76 & \textbf{12} & \textbf{174} & \textbf{52} & \textbf{30}\\
		\cline{5-8}	
		6 & 1,107 & 661 & 60 \\
		\cline{1-4}	
		\textbf{4-6} & \textbf{9,195} & \textbf{7,662} & \textbf{84} \\		              
		\cline{1-4}
	\end{tabular}
	\caption{The accuracy (A [$\%$]) of the PHOT classification given as the percentage of the number of galaxies in each SPEC class ($N_{spectro}$) to the subsample of $N_{spectro}$ galaxies which are found in the same PHOT class ($N_{photo}$).
	}             
	\label{table:eff}     
\end{table}

We conclude that reliable $z_{phot}$ ($\sigma \sim 0.03, \eta \sim 2\%$) allow for a highly effective identification of different galaxy classes at $z>0.5$ by the FEM unsupervised algorithm.
The resultant classification is almost identical with the classification based on $z_{spec}$, confirming that the FEM algorithm is a new robust tool, which can provide a detailed picture of galaxy types at $z\sim0.7$ also without spectroscopic information.
In general, it implies that for division into basic blue/green/red classes at $z\sim0.7$ the observed colours are sufficient and even incorrectly estimated redshifts do not affect the results significantly (especially the selection of red galaxies). 
However, for a finer division into subclasses, a correct redshift is necessary. 
To reproduce most subclasses the photometric accuracy as provided by our data is sufficient, except for blue dusty star-forming galaxies (SPEC class 8), where spectroscopic accuracy becomes crucial.
The proper separation of this class, as well as broad-line AGNs and stars may depend on the accuracy of the SED templates of these objects. 
Overall, we have found that the purely photometric classification is not strongly biased with respect to the spectroscopic one. 
This excellent agreement gives us confidence that future photometric surveys can produce robust galaxy classifications useful for galaxy evolution studies.

\section*{Acknowledgements}

VIPERS is based on observations collected at the European Southern Observatory, Cerro Paranal, Chile, using the Very Large Telescope under programs 182.A-0886 and partly 070.A-9007. 
Also based on observations obtained with MegaPrime/MegaCam, a joint project of CFHT and CEA/DAPNIA, at the Canada-France-Hawaii Telescope (CFHT), which is operated by the	National Research Council (NRC) of Canada, the Institut National des Sciences de l'Univers of the Centre National de la Recherche Scientifique (CNRS) of France, and the University of Hawaii. This work is based in part on data products produced at TERAPIX and the Canadian Astronomy Data Centre as part of the Canada-France-Hawaii Telescope Legacy Survey, a collaborative project of NRC and CNRS. 
We acknowledge the crucial contribution of the ESO staff for the management of service observations. In particular, we are deeply grateful to M. Hilker for his constant help and support of this program. Italian participation in VIPERS has been funded by INAF through PRIN 2008, 2010, and 2014 programs. LG and BRG acknowledge support of the European Research Council through the Darklight ERC Advanced Research Grant (\# 291521). OLF acknowledges support of the European Research Council through the EARLY ERC Advanced Research Grant (\# 268107). KM, JK, MS have been supported by the National Science Centre (grant UMO-2013/09/D/ST9/04030). MS also acknowledges financial support from UMO-2016/23/N/ST9/02963 and AP from UMO-2017/26/A/ST9/00756 by the National Science Centre . RT acknowledge financial support from the European Research Council under the European Community's Seventh Framework Programme (FP7/2007-2013)/ERC grant agreement n. 202686. EB, FM, and LM acknowledge the support from grants ASI-INAF I/023/12/0 and PRIN MIUR 2010-2011. LM also acknowledges financial support from PRIN INAF 2012.




\bibliographystyle{mnras}
\bibliography{vipers}

\begin{thebibliography}{}
\makeatletter
\relax
\def\mn@urlcharsother{\let\do\@makeother \do\$\do\&\do\#\do\^\do\_\do\%\do\~}
\def\mn@doi{\begingroup\mn@urlcharsother \@ifnextchar [ {\mn@doi@}
  {\mn@doi@[]}}
\def\mn@doi@[#1]#2{\def\@tempa{#1}\ifx\@tempa\@empty \href
  {http://dx.doi.org/#2} {doi:#2}\else \href {http://dx.doi.org/#2} {#1}\fi
  \endgroup}
\def\mn@eprint#1#2{\mn@eprint@#1:#2::\@nil}
\def\mn@eprint@arXiv#1{\href {http://arxiv.org/abs/#1} {{\tt arXiv:#1}}}
\def\mn@eprint@dblp#1{\href {http://dblp.uni-trier.de/rec/bibtex/#1.xml}
  {dblp:#1}}
\def\mn@eprint@#1:#2:#3:#4\@nil{\def\@tempa {#1}\def\@tempb {#2}\def\@tempc
  {#3}\ifx \@tempc \@empty \let \@tempc \@tempb \let \@tempb \@tempa \fi \ifx
  \@tempb \@empty \def\@tempb {arXiv}\fi \@ifundefined
  {mn@eprint@\@tempb}{\@tempb:\@tempc}{\expandafter \expandafter \csname
  mn@eprint@\@tempb\endcsname \expandafter{\@tempc}}}

\bibitem[\protect\citeauthoryear{{Arnouts} et~al.,}{{Arnouts}
  et~al.}{2013}]{arnouts}
{Arnouts} S.,  et~al., 2013, \mn@doi [\aap] {10.1051/0004-6361/201321768},
  \href {http://adsabs.harvard.edu/abs/2013A%26A...558A..67A} {558, A67}

\bibitem[\protect\citeauthoryear{{Balogh}, {Morris}, {Yee}, {Carlberg}  \&
  {Ellingson}}{{Balogh} et~al.}{1999}]{balogh1999}
{Balogh} M.~L.,  {Morris} S.~L.,  {Yee} H.~K.~C.,  {Carlberg} R.~G.,
  {Ellingson} E.,  1999, \mn@doi [\apj] {10.1086/308056}, \href
  {http://adsabs.harvard.edu/abs/1999ApJ...527...54B} {527, 54}

\bibitem[\protect\citeauthoryear{{Bilicki} et~al.,}{{Bilicki}
  et~al.}{2018}]{bilicki2018}
{Bilicki} M.,  et~al., 2018, \mn@doi [\aap] {10.1051/0004-6361/201731942},
  \href {http://adsabs.harvard.edu/abs/2018A%26A...616A..69B} {616, A69}

\bibitem[\protect\citeauthoryear{{Bouveyron} \& {Brunet}}{{Bouveyron} \&
  {Brunet}}{2011}]{Bou2011}
{Bouveyron} C.,  {Brunet} C.,  2011, preprint (\mn@eprint {} {1101.2374})

\bibitem[\protect\citeauthoryear{{Dom{\'{\i}}nguez S{\'a}nchez},
  {Huertas-Company}, {Bernardi}, {Tuccillo}  \& {Fischer}}{{Dom{\'{\i}}nguez
  S{\'a}nchez} et~al.}{2018}]{sanchez2018}
{Dom{\'{\i}}nguez S{\'a}nchez} H.,  {Huertas-Company} M.,  {Bernardi} M.,
  {Tuccillo} D.,   {Fischer} J.~L.,  2018, \mn@doi [\mnras]
  {10.1093/mnras/sty338}, \href
  {http://adsabs.harvard.edu/abs/2018MNRAS.476.3661D} {476, 3661}

\bibitem[\protect\citeauthoryear{{Fraix-Burnet}, {Thuillard}  \&
  {Chattopadhyay}}{{Fraix-Burnet} et~al.}{2015}]{FBD15}
{Fraix-Burnet} D.,  {Thuillard} M.,   {Chattopadhyay} A.~K.,  2015, \mn@doi
  [Frontiers in Astronomy and Space Sciences] {10.3389/fspas.2015.00003}, \href
  {http://adsabs.harvard.edu/abs/2015FrASS...2....3F} {2, 3}

\bibitem[\protect\citeauthoryear{{Fritz} et~al.,}{{Fritz}
  et~al.}{2014}]{fritz14}
{Fritz} A.,  et~al., 2014, \aap, \href
  {http://adsabs.harvard.edu/abs/2014arXiv1401.6137F} {563, A92}

\bibitem[\protect\citeauthoryear{{George} \& {Huerta}}{{George} \&
  {Huerta}}{2018}]{george2018}
{George} D.,  {Huerta} E.~A.,  2018, \mn@doi [\prd]
  {10.1103/PhysRevD.97.044039}, \href
  {http://adsabs.harvard.edu/abs/2018PhRvD..97d4039G} {97, 044039}

\bibitem[\protect\citeauthoryear{{Guzzo} et~al.,}{{Guzzo} et~al.}{2014}]{guzzo}
{Guzzo} L.,  et~al., 2014, \mn@doi [\aap] {10.1051/0004-6361/201321489}, \href
  {http://adsabs.harvard.edu/abs/2014A%26A...566A.108G} {566, A108}

\bibitem[\protect\citeauthoryear{{Haines} et~al.,}{{Haines}
  et~al.}{2017}]{haines16}
{Haines} C.~P.,  et~al., 2017, \mn@doi [\aap] {10.1051/0004-6361/201630118},
  \href {http://adsabs.harvard.edu/abs/2017A%26A...605A...4H} {605, A4}

\bibitem[\protect\citeauthoryear{{Hocking}, {Geach}, {Sun}  \&
  {Davey}}{{Hocking} et~al.}{2018}]{Hocking2018}
{Hocking} A.,  {Geach} J.~E.,  {Sun} Y.,   {Davey} N.,  2018, \mn@doi [\mnras]
  {10.1093/mnras/stx2351}, \href
  {http://adsabs.harvard.edu/abs/2018MNRAS.473.1108H} {473, 1108}

\bibitem[\protect\citeauthoryear{{Hudelot} et~al.,}{{Hudelot}
  et~al.}{2012}]{Hudelot2012}
{Hudelot} P.,  et~al., 2012, VizieR Online Data Catalog, \href
  {http://adsabs.harvard.edu/abs/2012yCat.2317....0H} {2317}

\bibitem[\protect\citeauthoryear{{Ilbert}, {Arnouts}, {McCracken}
  et~al.}{{Ilbert} et~al.}{2006}]{ilbert2006}
{Ilbert} O.,  {Arnouts} S.,  {McCracken} H.~J.,   et~al., 2006, A\&A, 457, 841

\bibitem[\protect\citeauthoryear{{Ivezic} et~al.,}{{Ivezic}
  et~al.}{2008}]{ivezic2008}
{Ivezic} Z.,  et~al., 2008, preprint, \href
  {http://adsabs.harvard.edu/abs/2008arXiv0805.2366I} {} (\mn@eprint {arXiv}
  {0805.2366})

\bibitem[\protect\citeauthoryear{{Jarvis} et~al.,}{{Jarvis}
  et~al.}{2013}]{Jarvis2013}
{Jarvis} M.~J.,  et~al., 2013, \mn@doi [\mnras] {10.1093/mnras/sts118}, \href
  {http://adsabs.harvard.edu/abs/2013MNRAS.428.1281J} {428, 1281}

\bibitem[\protect\citeauthoryear{{Kauffmann} et~al.,}{{Kauffmann}
  et~al.}{2003}]{kauffmann03}
{Kauffmann} G.,  et~al., 2003, \mn@doi [\mnras]
  {10.1046/j.1365-8711.2003.06291.x}, \href
  {http://adsabs.harvard.edu/abs/2003MNRAS.341...33K} {341, 33}

\bibitem[\protect\citeauthoryear{{Krakowski}, {Ma{\l}ek}, {Bilicki}, {Pollo},
  {Kurcz}  \& {Krupa}}{{Krakowski} et~al.}{2016}]{Krakowski16}
{Krakowski} T.,  {Ma{\l}ek} K.,  {Bilicki} M.,  {Pollo} A.,  {Kurcz} A.,
  {Krupa} M.,  2016, \mn@doi [\aap] {10.1051/0004-6361/201629165}, \href
  {http://adsabs.harvard.edu/abs/2016A%26A...596A..39K} {596, A39}

\bibitem[\protect\citeauthoryear{{Krywult} et~al.,}{{Krywult}
  et~al.}{2017}]{krywult}
{Krywult} J.,  et~al., 2017, \mn@doi [\aap] {10.1051/0004-6361/201628953},
  \href {http://adsabs.harvard.edu/abs/2017A%26A...598A.120K} {598, A120}

\bibitem[\protect\citeauthoryear{{Kurcz}, {Bilicki}, {Solarz}, {Krupa}, {Pollo}
   \& {Ma{\l}ek}}{{Kurcz} et~al.}{2016}]{Kurcz16}
{Kurcz} A.,  {Bilicki} M.,  {Solarz} A.,  {Krupa} M.,  {Pollo} A.,   {Ma{\l}ek}
  K.,  2016, \mn@doi [\aap] {10.1051/0004-6361/201628142}, \href
  {http://adsabs.harvard.edu/abs/2016A%26A...592A..25K} {592, A25}

\bibitem[\protect\citeauthoryear{{LSST Science Collaboration} et~al.,}{{LSST
  Science Collaboration} et~al.}{2009}]{lsst2009}
{LSST Science Collaboration} et~al., 2009, preprint, \href
  {http://adsabs.harvard.edu/abs/2009arXiv0912.0201L} {} (\mn@eprint {arXiv}
  {0912.0201})

\bibitem[\protect\citeauthoryear{{Laureijs} et~al.,}{{Laureijs}
  et~al.}{2011}]{euclid2011}
{Laureijs} R.,  et~al., 2011, preprint, \href
  {http://adsabs.harvard.edu/abs/2011arXiv1110.3193L} {} (\mn@eprint {arXiv}
  {1110.3193})

\bibitem[\protect\citeauthoryear{{Le F{\`e}vre} et~al.,}{{Le F{\`e}vre}
  et~al.}{2003}]{lefevre03}
{Le F{\`e}vre} O.,  et~al., 2003, in {Iye} M.,  {Moorwood} A.~F.~M.,  eds,
  Society of Photo-Optical Instrumentation Engineers (SPIE) Conference Series
  Vol. 4841, Society of Photo-Optical Instrumentation Engineers (SPIE)
  Conference Series. pp 1670--1681, \mn@doi{10.1117/12.460959}

\bibitem[\protect\citeauthoryear{{Ma{\l}ek} et~al.,}{{Ma{\l}ek}
  et~al.}{2013}]{malek13}
{Ma{\l}ek} K.,  et~al., 2013, \mn@doi [\aap] {10.1051/0004-6361/201321447},
  \href {http://adsabs.harvard.edu/abs/2013A%26A...557A..16M} {557, A16}

\bibitem[\protect\citeauthoryear{{Marchetti} et~al.,}{{Marchetti}
  et~al.}{2012}]{marchetti12}
{Marchetti} A.,  et~al., 2012, \mn@doi [\mnras] {10.1093/mnras/sts132}, \href
  {http://adsabs.harvard.edu/abs/2012MNRAS.tmp..107M} {p.~107}

\bibitem[\protect\citeauthoryear{{Martin} et~al.,}{{Martin}
  et~al.}{2005}]{Martin2005}
{Martin} D.~C.,  et~al., 2005, \mn@doi [\apjl] {10.1086/426387}, \href
  {http://adsabs.harvard.edu/abs/2005ApJ...619L...1M} {619, L1}

\bibitem[\protect\citeauthoryear{{Moresco} et~al.,}{{Moresco}
  et~al.}{2013}]{moresco13}
{Moresco} M.,  et~al., 2013, \mn@doi [\aap] {10.1051/0004-6361/201321797},
  \href {http://adsabs.harvard.edu/abs/2013A%26A...558A..61M} {558, A61}

\bibitem[\protect\citeauthoryear{{Moutard} et~al.,}{{Moutard}
  et~al.}{2016a}]{moutard16a}
{Moutard} T.,  et~al., 2016a, \mn@doi [\aap] {10.1051/0004-6361/201527945},
  \href {http://adsabs.harvard.edu/abs/2016A\%26A...590A.102M} {590, A102}

\bibitem[\protect\citeauthoryear{{Moutard} et~al.,}{{Moutard}
  et~al.}{2016b}]{moutard16b}
{Moutard} T.,  et~al., 2016b, \mn@doi [\aap] {10.1051/0004-6361/201527294},
  \href {http://adsabs.harvard.edu/abs/2016A\%26A...590A.103M} {590, A103}

\bibitem[\protect\citeauthoryear{{Parks}, {Prochaska}, {Dong}  \&
  {Cai}}{{Parks} et~al.}{2018}]{parks2018}
{Parks} D.,  {Prochaska} J.~X.,  {Dong} S.,   {Cai} Z.,  2018, \mn@doi [\mnras]
  {10.1093/mnras/sty196}, \href
  {http://adsabs.harvard.edu/abs/2018MNRAS.476.1151P} {476, 1151}

\bibitem[\protect\citeauthoryear{{Peth} et~al.,}{{Peth}
  et~al.}{2016}]{Peth2016}
{Peth} M.~A.,  et~al., 2016, \mn@doi [\mnras] {10.1093/mnras/stw252}, \href
  {http://adsabs.harvard.edu/abs/2016MNRAS.458..963P} {458, 963}

\bibitem[\protect\citeauthoryear{{Renzini}}{{Renzini}}{2006}]{renzini}
{Renzini} A.,  2006, \mn@doi [\araa] {10.1146/annurev.astro.44.051905.092450},
  \href {http://adsabs.harvard.edu/abs/2006ARA%26A..44..141R} {44, 141}

\bibitem[\protect\citeauthoryear{{S{\'a}nchez Almeida}, {Aguerri},
  {Mu{\~n}oz-Tu{\~n}{\'o}n}  \& {de Vicente}}{{S{\'a}nchez Almeida}
  et~al.}{2010}]{SDSS_class_2010}
{S{\'a}nchez Almeida} J.,  {Aguerri} J.~A.~L.,  {Mu{\~n}oz-Tu{\~n}{\'o}n} C.,
  {de Vicente} A.,  2010, \mn@doi [\apj] {10.1088/0004-637X/714/1/487}, \href
  {http://adsabs.harvard.edu/abs/2010ApJ...714..487S} {714, 487}

\bibitem[\protect\citeauthoryear{{Scodeggio} et~al.,}{{Scodeggio}
  et~al.}{2018}]{scodeggio16}
{Scodeggio} M.,  et~al., 2018, \mn@doi [\aap] {10.1051/0004-6361/201630114},
  \href {http://adsabs.harvard.edu/abs/2018A%26A...609A..84S} {609, A84}

\bibitem[\protect\citeauthoryear{{Shallue} \& {Vanderburg}}{{Shallue} \&
  {Vanderburg}}{2018}]{shallue2018}
{Shallue} C.~J.,  {Vanderburg} A.,  2018, \mn@doi [\aj]
  {10.3847/1538-3881/aa9e09}, \href
  {http://adsabs.harvard.edu/abs/2018AJ....155...94S} {155, 94}

\bibitem[\protect\citeauthoryear{{Siudek} et~al.,}{{Siudek}
  et~al.}{2017}]{siudek17}
{Siudek} M.,  et~al., 2017, \mn@doi [\aap] {10.1051/0004-6361/201628951}, \href
  {http://adsabs.harvard.edu/abs/2017A%26A...597A.107S} {597, A107}

\bibitem[\protect\citeauthoryear{{Siudek} et~al.,}{{Siudek}
  et~al.}{2018}]{siudek18}
{Siudek} M.,  et~al., 2018, \mn@doi [\aap] {10.1051/0004-6361/201832784}, \href
  {http://adsabs.harvard.edu/abs/2018A%26A...617A..70S} {617, A70}

\bibitem[\protect\citeauthoryear{{Solarz} et~al.,}{{Solarz}
  et~al.}{2012}]{Solarz2012}
{Solarz} A.,  et~al., 2012, \mn@doi [\aap] {10.1051/0004-6361/201118108}, \href
  {http://adsabs.harvard.edu/abs/2012A%26A...541A..50S} {541, A50}

\bibitem[\protect\citeauthoryear{{Solarz}, {Bilicki}, {Gromadzki}, {Pollo},
  {Durkalec}  \& {Wypych}}{{Solarz} et~al.}{2017}]{Solarz2017}
{Solarz} A.,  {Bilicki} M.,  {Gromadzki} M.,  {Pollo} A.,  {Durkalec} A.,
  {Wypych} M.,  2017, \mn@doi [\aap] {10.1051/0004-6361/201730968}, \href
  {http://adsabs.harvard.edu/abs/2017A%26A...606A..39S} {606, A39}

\makeatother
\end{thebibliography}

%
$^{1}$Institut de F'isica d'Altes Energies (IFAE), The Barcelona Institute of Science and Technology, 08193 Bellaterra (Barcelona), Spain\\
$^{2}$Centre for Theoretical Physics, Al. Lotnikow 32/46, 02-668 Warsaw, Poland\\
$^{3}$National Centre for Nuclear Research, ul. Hoza 69, 00-681 Warszawa, Poland\\
$^{4}$Astronomical Observatory of the Jagiellonian University, Orla 171, 30-001 Cracow, Poland\\
$^{5}$ INAF - Osservatorio Astronomico di Brera, Via Brera 28, 20122 Milano
--  via E. Bianchi 46, 23807 Merate, Italy\\
$^{6}$Universit\`{a} degli Studi di Milano, via G. Celoria 16, 20133 Milano, Italy\\
$^{7}$INAF - Istituto di Astrofisica Spaziale e Fisica Cosmica Milano, via Bassini 15, 20133 Milano, Italy\\
$^{8}$Department of Astronomy \& Physics, Saint Mary's University, 923 Robie Street, Halifax, Nova Scotia, B3H 3C3, Canada\\
$^{9}$Aix Marseille Univ, CNRS, LAM, Laboratoire d'Astrophysique de
Marseille, Marseille, France  \\
$^{10}$INAF - Osservatorio di Astrofisica e Scienza dello Spazio di Bologna,	via Gobetti 93/3,	40129 Bologna - Italy\\
$^{11}$INAF - Osservatorio Astrofisico di Torino, 10025 Pino Torinese, Italy \\
$^{12}$Laboratoire Lagrange, UMR7293, Universit\'e de Nice Sophia Antipolis, CNRS, Observatoire de la C\^ote d'Azur, 06300 Nice, France\\
$^{13}$Dipartimento di Fisica e Astronomia - Alma Mater Studiorum Universit\`{a} di Bologna, via Gobetti 93/2, I-40129, Bologna, Italy\\
$^{14}$Institute of Physics, Jan Kochanowski University, ul. Swietokrzyska 15, 25-406 Kielce, Poland\\
$^{15}$INFN, Sezione di Bologna, viale Berti Pichat 6/2, I-40127, Bologna, Italy\\
$^{16}$IRAP, Universit\`{e} de Toulouse, CNRS, UPS, Toulouse, France\\
$^{17}$IRAP,  9 av. du colonel Roche, BP 44346, F-31028 Toulouse cedex 4, France\\ 
$^{18}$School of Physics and Astronomy, University of St Andrews, St Andrews KY16 9SS, UK\\
$^{19}$INAF - Istituto di Radioastronomia, via Gobetti 101, I-40129, Bologna, Italy\\
$^{20}$Canada-France-Hawaii Telescope, 65--1238 Mamalahoa Highway, Kamuela, HI 96743, USA\\
$^{21}$Aix Marseille Univ, Univ Toulon, CNRS, CPT, Marseille, France\\
$^{22}$Dipartimento di Matematica e Fisica, Universit\`{a} degli Studi Roma Tre, via della Vasca Navale 84, 00146 Roma, Italy\\
$^{23}$INFN, Sezione di Roma Tre, via della Vasca Navale 84, I-00146 Roma, Italy\\
$^{24}$INAF - Osservatorio Astronomico di Roma, via Frascati 33, I-00040 Monte Porzio Catone (RM), Italy\\
$^{25}$Department of Astronomy, University of Geneva, ch. d'Ecogia 16, 1290 Versoix, Switzerland\\
$^{26}$INAF - Osservatorio Astronomico di Trieste, via G. B. Tiepolo 11, 34143 Trieste, Italy\\
$^{27}$Division of Particle and Astrophysical Science, Nagoya University, Furo-cho, Chikusa-ku, 464-8602 Nagoya, Japan\\

\bsp	
\label{lastpage}
\end{document}